# Magnetic Exchange Fields and Domain Wall Superconductivity at an All-Oxide Superconductor-Ferromagnet Insulator Interface

S. Komori,[1] A. Di Bernardo,[1] A. I. Buzdin,[1, 2] M. G. Blamire,[1] and J. W. A. Robinson[1, *]

jjr33@cam.ac.uk

[1]*Department of Materials Science and Metallurgy, University of Cambridge, 27 Charles Babbage Road, Cambridge CB3 0FS, United Kingdom*

[2]*University Bordeaux, LOMA UMR-CNRS 5798, F-33405 Talence Cedex, France*

   At a superconductor-ferromagnet (*S*/*F*) interface, the *F* layer can introduce a magnetic exchange field within the S layer, which acts to locally spin split the superconducting density of states. The effect of magnetic exchange fields on superconductivity has been thoroughly explored at *S*-ferromagnet insulator (*S*/FI) interfaces for isotropic *s*-wave *S* and a thickness that is smaller than the superconducting coherence length. Here we report a magnetic exchange field effect at an all-oxide *S*/FI interface involving the anisotropic *d*-wave high temperature superconductor praseodymium cerium copper oxide (PCCO) and the FI praseodymium calcium manganese oxide (PCMO). The magnetic exchange field in PCCO, detected via magnetotransport measurements through the superconducting transition, is localized to the PCCO/PCMO interface with an average magnitude that depends on the presence or absence of magnetic domain walls in PCMO. The results are promising for the development of all-oxide superconducting spintronic devices involving unconventional pairing and high temperature superconductors.

The proximity effects at a superconductor-ferromagnet metal ($S/F$) interface describe the leakage of superconductivity from $S$ into $F$ and the penetration of a magnetic exchange field (MEF) from $F$ into $S$ [1]. MEFs within $F$ act differentially on the antiparallel spins of the singlet Cooper pairs, which introduces oscillating components to the superconducting order parameter in $F$ [2–5]. For $S/F$ bilayers, this results in a nonmonotonic critical temperature ($T_c$) on $F$ layer thickness [6–9] and in $S/F/S$ Josephson junctions, to a modulation of the critical current between 0 and π states as a function of $F$ thickness [2,10–19].

The coherence length ($\xi_F$) of spin-singlet pairs in a $F$ metal is short ranged and less than 3 nm in Co [13–16], Fe [16,17,20], and Ni [13,15,16]. An inhomogeneous MEF at a $S/F$ interface can convert spin-singlet pairs into a spinaligned triplet state [21–24] and so increase $\xi_F$ to tens of nanometers in $F$ metals [25–34] and potentially to hundreds of nanometers in half-metallic $F$s [25,29,35–40]. Since spin-aligned triplet pairs are therefore sensitive to the micromagnetic state [36,41–46] of $S/F$ heterostructures, their recent discovery and control in $S/F$ devices has led the new area of superconducting spintronics [21].

On the $S$ side of a $S/F$ interface, the MEF penetrates over a distance of less than the superconducting coherence length $\xi_s$ and acts to spin split the superconducting density of states (DOS) [47–49]. The effect of a MEF on superconductivity is well understood at $S$-ferromagnet insulator ($S$/FI) interfaces [50–53] involving isotropic $s$-wave $S$—e.g., tunneling spectroscopy on superconducting Al with a thickness below $\xi_s$ on EuS (FI) has demonstrated spin splitting in Al that is equivalent to an external magnetic field of several Tesla [47,54,55]. Similar results are also reported in NbN/GdN [50], Nb/GdN [52], and In/Fe$_3$O$_4$ [53]. However, the MEF effect at a $S$/FI interface involving anisotropic $d$-wave high temperature superconductor (HTS) oxides has hardly been explored either experimentally or theoretically. This might be because the $c$-axis coherence length ($\xi_c$) of HTS oxides is subnanometer [e.g., $\xi_c \approx 0.3$–0.4 nm in hole-doped YBa$_2$Cu$_3$O$_7$ (YBCO) [56,57] and $\xi_c \approx 1$ nm in electron-doped Pr$_{1.85}$Ce$_{0.15}$CuO$_4$ (PCCO) [58,59] at absolute zero] meaning a MEF is challenging to detect. Since superconductivity is quenched in HTS oxide thin films even at thicknesses larger than $\xi_c$ [60,61] due to factors including, e.g., strain [62,63], oxygen deficiencies [64,65], and structural defects [64–66], tunneling spectroscopy has to be performed on much thicker films where the interface DOS is masked by the bulk superconductivity. Alternatively, MEFs can be probed via magnetotransport measurements around $T_c$. However, measurements on YBCO/La$_{0.7}$Ca$_{0.3}$MnO$_3$ (LCMO) show results dominated by stray fields from Bloch domain walls in LCMO, which suppress $T_c$ [67,68], and in LCMO/YBCO/LCMO pseudo-spin-valves, the $T_c$ is reduced by spin accumulation in YBCO [69–72].

In this Letter, we report PCCO-thickness-dependent magnetotransport measurements around the $T_c$ of PCCO proximity coupled to the FI Pr$_{0.8}$Ca$_{0.2}$MnO$_3$ (PCMO). By sweeping in-plane magnetic fields around the coercivity of the PCMO as a function of PCCO layer thickness, we are able to correlate shifts in $T_c$ to the localized MEF in PCCO at the PCCO/PCMO interface. A suppression of $T_c$ due to stray fields from Bloch domains walls in PCMO or nonequilibrium accumulation of spin-polarized quasiparticles in PCCO are ruled out.

PCCO/PCMO and PCMO/PCCO/PCMO films are grown on (001)-oriented single crystal SrTiO$_3$ (STO) by pulsed laser deposition (KrF laser; wavelength λ = 248 nm) at a growth temperature of 780 °C in flowing $N_2O$ at 120 mTorr with a beam energy density of 1.5 J cm$^{-2}$ and pulse frequency of 7 Hz for PCCO and 4 Hz for PCMO. The films are postannealed *in situ* at 720 °C for several minutes in 0.1 mTorr of $N_2O$ to optimize the superconductivity of the PCCO and subsequently cooled at a rate of 5 °Cmin$^{-1}$. In-plane electrical resistance ($R$) measurements using a current of 100 $\mu$A were performed in a cryogen-free pulse tube system using a four-terminal electrical setup with 20-nm-thick Au contacts on the films. R was measured as a function of an in-plane magnetic field ($H$), directed parallel to the applied current, and temperature ($T$) across $T_c$. Care was taken to ensure that the current had no effect on $T_c$.

In Fig. 1(a), we have plotted high resolution x-ray diffraction data from a PCCO/PCMO bilayer, which confirm (001) *c*-axis growth of PCMO on STO with rocking curves on the pseudocubic (001)$_c$, (002)$_c$ , and (003)$_c$ Bragg peaks, showing full width at half maximum(FWHM) values of 0.14 °, 0.23 °, and 0.24 ° [Fig. 1(b)]. The PCCO is (001) textured, but contains a component from the (110) phase with FWHM values of 1.2 °–1.3 ° for both orientations [Figs. 1(c) and 1(d)]. The volume fraction of the (110) phase is estimated to be 6.2% from a comparison with x-ray powder diffraction data [73]. The average *c*-axis lattice parameters determined from multiple high angle diffraction peaks for PCCO and PCMO are 1.218 and 0.769 nm, respectively, consistent with x-ray powder diffraction on PCCO [73,74] and PCMO [75]. Figure S1 in the Supplemental Material [76] shows the topography of a PCMO/PCCO/PCMO (53/26/106 nm) trilayer, from which we estimate a root-mean-square roughness of less than 1 nm over an area of 25 $\mu$m$^2$.

We first discuss $R(H)$ results taken from PCCO/PCMO bilayers in the superconducting transition. Figure 2(a) shows $R(H)$ at 4.5 K for PCCO/PCMO (26/106 nm) with an onset critical temperature ($T_{c,\text{onset}}$) of 13.5 K, alongside the corresponding in-plane magnetic moment vs magnetic field $m(H)$ loop and $R(T)$. The $U$-shaped background in $R(T)$ is due to field suppression of superconductivity; around the coercive field ($H_c$) of PCMO (±0.09 T), $R$ decreases relative to the background, which translates to an increase in $T_c$. As the PCMO magnetic moment approaches saturation (i.e., $H > H_c$), $R$ increases, meaning that $T_c$ decreases. Similar results were obtained for PCMO/PCCO/PCMO (53/94/106 nm) trilayers [Fig. 2(b)]. These results suggest that, if out-of-plane Bloch domain walls exist in PCMO, they have no measurable effect on the superconductivity of PCCO, as such fields would suppress $T_c$ at $H_c$ where the domain wall density is maximum (e.g. see [84]). From here on, we refer to shifts in $R$ between magnetized ($R_{H>Hc}$) and demagnetized ($R_{Hc}$) states of PCMO as $\Delta R = R_{H>Hc} - R_{Hc}$, as shown in Fig. 2(a). We note that, for all PCCO/PCMO and PCMO/PCCO/PCMO samples, $\Delta R = 0$ for $T > T_{c,\text{onset}}$, thus ruling out normal-state magnetoresistance as an explanation for $\Delta R$ in the superconducting transition [see inset to Fig. 2(b)].

To further eliminate any potential influence of stray fields on $R(H)$, we fabricated a PCMO/PCO/PCCO/PCO/PCMO (53/3/94/3/106 nm) control sample in which the insulating layer of $Pr_2CuO_4$ (PCO) is nonmagnetic and blocks proximity coupling between PCCO and PCMO. As shown in Fig. 2(c), $\Delta R = 0$ around $H_c$, meaning that any stray fields, which should still be present if out-of-plane Bloch domain walls exist, have negligible effect on superconductivity.

Note that the $T_c$ of PCCO in all multilayers investigated is reduced from the bulk value of 22 K [85]. In the Supplemental Material [76], we investigate in detail the dependence of $T_c$ for PCCO as a function of PCCO layer thickness ($d_{PCCO}$) for PCCO($d_{PCCO}$)/STO, PCCO($d_{PCCO}$)/PCMO(106 nm)/STO, and PCCO($d_{PCCO}$)/PCO(3 nm)/PCMO(106 nm)/STO structures. For PCCO($d_{PCCO}$)/STO, a weaker suppression of $T_c$ with decreasing $d_{PCCO}$ is observed compared to PCCO($d_{PCCO}$)/PCMO(106 nm)/STO and PCCO($d_{PCCO}$)/PCO(3 nm)/PCMO(106 nm)/STO structures. This is explained on the basis of two competing effects: in-plane compressive strain in PCCO which enhances $T_c$ [62,63,77,78,83] and a reduced net removal of oxygen from the apical positions ($O_a$) of PCCO during the annealing process which suppresses $T_c$ [66,83]. During annealing, PCCO on STO and PCCO on PCMO lose $O_a$, but for PCCO on PCMO, the net loss will be lower since $O^{2-}$ ions will diffuse from PCMO into PCCO, and hence a greater suppression of $T_c$ is expected than for PCCO on STO. Furthermore, in-plane compressive strain in PCCO on STO (−0.58%) is higher than for PCCO on PCMO (−0.08%), which enhances the $T_c$ of PCCO on STO compared to PCCO on PCMO.

We now show that the decrease in $T_c$ at $H_c$ is related to an interface MEF, which acts on the superconductivity in PCCO within a distance of $\xi_c(T)$ of the PCCO/PCMO interface. For $H > H_c$, the PCMO is magnetically saturated, and so the Cooper pairs in PCCO experience a spatially uniform MEF through an exchange interaction [86–89] with the electrons in Mn $3d\,e_g^1$ band at the interface of PCMO, which introduces a spin splitting in the superconducting DOS and an interface suppression of $T_c$ (see related works in [47,48,50,54,55]). However, around $H_c$, the magnetization in PCMO is inhomogeneous with magnetic domains pointing in different directions with the domain wall density therefore maximized.

The shift in $T_c$ between magnetized and demagnetized states in PCMO depends on the ratio $\alpha = \xi_{ab}(0)/d_w$, where $\xi_{ab}(0)$ is the in-plane coherence length in PCCO (at absolute zero) and $d_w$ is the domain wall width in PCMO, via the relation [1,90,91]

$$\frac{\Delta T_c}{T_c} \approx \frac{\pi^6 \alpha^2}{36}\left(\frac{h}{2\pi T_{c0}}\right)^4 \approx \frac{\alpha^2}{2}\frac{(T_{c0}-T_c)^2}{T_{c0}^2}, \qquad (1)$$

in the limit of $\alpha < 1$.

Here, $h$ is the magnitude of the MEF in $S$ and $T_{c0}$ is the critical temperature of the equivalent $S$ layer isolated from the FI—consequently, $\Delta R$ increases with $\alpha$. In manganites, $d_w$ is ≈20 nm [92] at temperatures relevant to our experiment. Values of $\xi_{ab}(0)$ and $\xi_c(0)$ are estimated from measurements of the upper critical field ($H_{c2}$) of PCCO: in Fig. 3(a), we have plotted the $T$ dependence of $H_{c2}$ for a PCMO/PCCO/PCMO (53/94/106 nm) trilayer—the $T = 0$ upper critical field out of plane [$H_{c2}^c(0)$] and in plane [$H_{c2}^{ab}(0)$] are 4.9 and 48 T, respectively. $H_{c2}^c$ was measured directly down to 4 K, while $H_{c2}^{ab}$ was estimated from Werthamer-Helfand-Hohenberg theory [93] to be $H_{c2}(0) = -0.69T_c(dH_{c2}/dT)|_{T_c}$. From $H_{c2}^c = \Phi_0/2\pi\xi_{ab}^2$ and $H_{c2}^{ab} = \Phi_0/2\pi\xi_{ab}\xi_c$, we estimate $\xi_c(0) = 0.83$ nm and $\xi_{ab}(0) = 8.1$ nm. Therefore, $\alpha$ for PCCO is 0.4, suggesting that at $H_c$ the Cooper pairs in PCCO above PCMO domain walls experience a reduced MEF and hence the local critical temperature is enhanced relative to the $T_c$ of PCCO above a uniformly magnetized domain, as illustrated in Fig. 3(b).

We now estimate $\Delta R$ due to domain wall superconductivity in PCCO. Taking the example of Fig. 2(b), we see that $\Delta R$ reaches a maximum value at $T = 16$ K with the PCCO close to the superconducting percolation threshold, where roughly 1/3 of the PCCO is superconducting [94]. Assuming that $\xi_{ab}$ follows $\xi_{ab}(T) \approx \xi_{ab}(0)/\sqrt{1 - \dfrac{T}{T_{c,\text{onset}}}}$, $\xi_{ab} \approx d_w \approx 20$ nm is obtained from $T_{c,\text{onset}} =$ 18 K [see $R(T)$ curve in Fig. 2(b)] and $\xi_{ab}(0) = 8.1$ nm. In this condition, the domain walls generate an additional superconducting region near the interface of volume of $V_s \approx d_w \xi_c(T) L_y$, where $L_y$ is the magnetic domain length. Because of the anisotropy of PCCO ($\xi_c << \xi_{ab} \approx d_w$), this region is elongated in the $ab$ plane. The domain wall superconductivity produces a strong perturbation on the current lines in PCCO at distances of the order of $\xi_{ab}(T)$ along the $c$ axis from the PCCO/PCMO interface, which increases the influence of the domain wall superconductivity on $\Delta R$ [see sketch in Fig. 3(b)]. Namely, the effective resistivity of the nonsuperconducting region within a distance $\xi_{ab}(T)$ from the domain wall is reduced due to the emergence of the domain wall superconductivity. The conductance of the PCCO will therefore be enhanced in regions across and along domain walls, with current filaments along the domain walls contributing most to the enhancement of the conductivity.

Therefore, $\Delta R/R_N$ (where $R_N$ is the normal-state resistance) can be estimated from the volume ratio of the region with reduced resistivity due to the presence of the domain wall superconductivity to the region above a uniformly magnetized domain

$$\frac{\Delta R}{R_N} = \frac{1}{3} \frac{\xi_{ab}^2(T)}{L d_{PCCO}}, \qquad (2)$$

where $L$ is the distance between magnetic domain walls. Taking $\Delta R/R_N \approx 2\%$ [inset of Fig. 2(b)], $d_w \approx$ 20 nm [95,96], $L \approx 100$ nm [95,96], and $d_{PCCO} = 47$ nm (half of the actual thickness), $\xi_{ab}(T) \approx 20$ nm which is consistent with our earlier assumption – i.e. $\xi_{ab}(T) \approx \frac{\xi_{ab}(0)}{\sqrt{1 - \frac{T}{T_{c,onset}}}} \approx d_w$ at $T$ = 18 K.

The difference between the local $T_c$ of PCCO above a domain wall ($T_{c,DW}$) versus above a magnetized region ($T_{c,MEF}$) of PCMO should be independent of $d_{PCCO}$ since $\alpha = \xi_{ab}(0)/d_w$ does not depend strongly on $d_{pcco}$ and the MEF acts within a fixed distance of $\xi_c(T)$ of the PCCO/PCMO interface. To estimate $\Delta T_c = T_{c,DW} - T_{c,MEF}$, we have calculated the effective change in $T_c$ (denoted $\Delta T_c'$) of PCCO by comparing $\Delta R$ from $R(T)$ to the slope of the superconducting transition. By plotting $\Delta T_c'$ as a function of $d_{PCCO}$ [Fig. 3(c)] and extrapolating $\Delta T_c'$ at $d_{PCCO} = \xi_c(T)$, we estimate $\Delta T_c$ to be larger than 50 mK and potentially of the order 1 K. Note that, although $\Delta T_c'$ [Fig. 3(c)] is inversely proportional to $d_{PCCO}$, this does not mean de Gennes theory [97] applies to PCCO/PCMO as the interface $\Delta T_c$ is independent of $d_{PCCO}$ —this behavior is therefore related to the conductance dependence of the nonsuperconducting regions of PCCO above a domain wall to $d_{PCCO}$.

Finally, we investigated $R(H)$ in a YBCO/PCMO (7/106 nm) bilayer [76] in which $\xi_c(0)$ is only 1–2 nm [56], meaning $\alpha$ for YBCO is 0.05–0.10 with $d_w$ = 20 nm [92] and $R(H)$ for a PCCO/LCMO (28/106 nm) bilayer [76]. For YBCO/PCMO, we do not observe switching in $R(H)$ at $H_c$, strongly supporting our claim that $\alpha = \xi_{ab}(0)/d_w$ should be close to unity in order to observe measurable shifts in $T_c$ due to a MEF. For PCCO/LCMO, we observe the opposite behavior to PCCO/PCMO with $R$ increasing at $H_c$, suggesting a spin-accumulation driven suppression of Tc in conjunction with stray fields from the LCMO.

In conclusion, we have demonstrated a MEF effect at an all-oxide PCCO/PCMO ($S$/FI) interface in which stray fields or spin accumulation have no detectable effect on the superconductivity. A MEF effect and therefore spin splitting in a HTS oxide is an important development for the fields of superconducting spintronics and oxide interfaces. The higher operating temperatures of HTS oxides over metallic-based superconductors in conjunction with their anisotropic pairing symmetries, could lead to novel forms of spin-dependent and thermal transport properties and so greatly extend the work on metallic-based systems—e.g., superconducting spin-polarized quasiparticle transport [98–100], $S$/FI thermoelectric effects [101,102], and FI/$S$/FI spin valves [51,52]. More exotically, an interface MEF could couple to surface (spin-polarized) bound states [103,104], which exist on anisotropic HTS oxides, as well as to spin fluctuations at the surface of the HTS, and so potentially influence the pairing mechanism itself in these materials.

The data sets relating to the figures in this paper are available for access in Ref. [105].

S. K., M. G. B., A. D. B., and J. W. A. R. acknowledge funding from the EPSRC through International Network and Programme Grants (No. EP/P026311/1 and No. EP/N017242/1). J. W. A. R. acknowledges funding from the Royal Society through a University Research Fellowship and with M. G. B. and A. I. B., funding from the Leverhulme Trust and EU Network COST CA16218 (NANOCOHYBRI). A. D. B. acknowledges funding from St. John's College, Cambridge. S. K. was supported by Grant-in-Aid for JSPS Research Fellows (No. 15J07623).

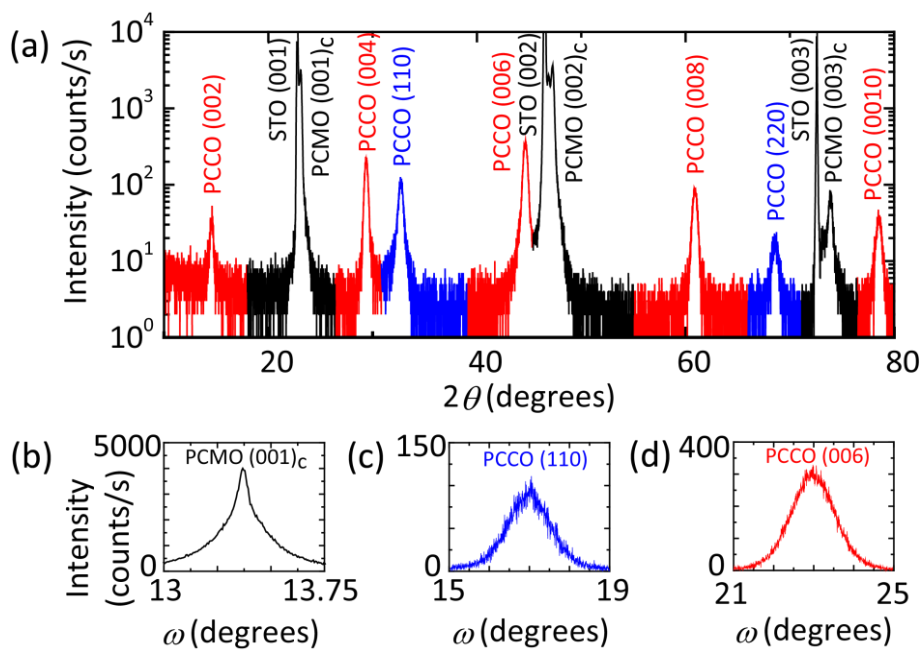

**FIG. 1.** (a) High angle x-ray diffraction from PCCO(26 nm)/PCMO(106 nm)/STO and corresponding (b)–(d) Rocking curves on the PCMO (001)$_c$, PCCO (004), PCCO(110), and PCCO (006) peaks showing FWHM values of 0.14°, 1.20°, and 1.32°, respectively.

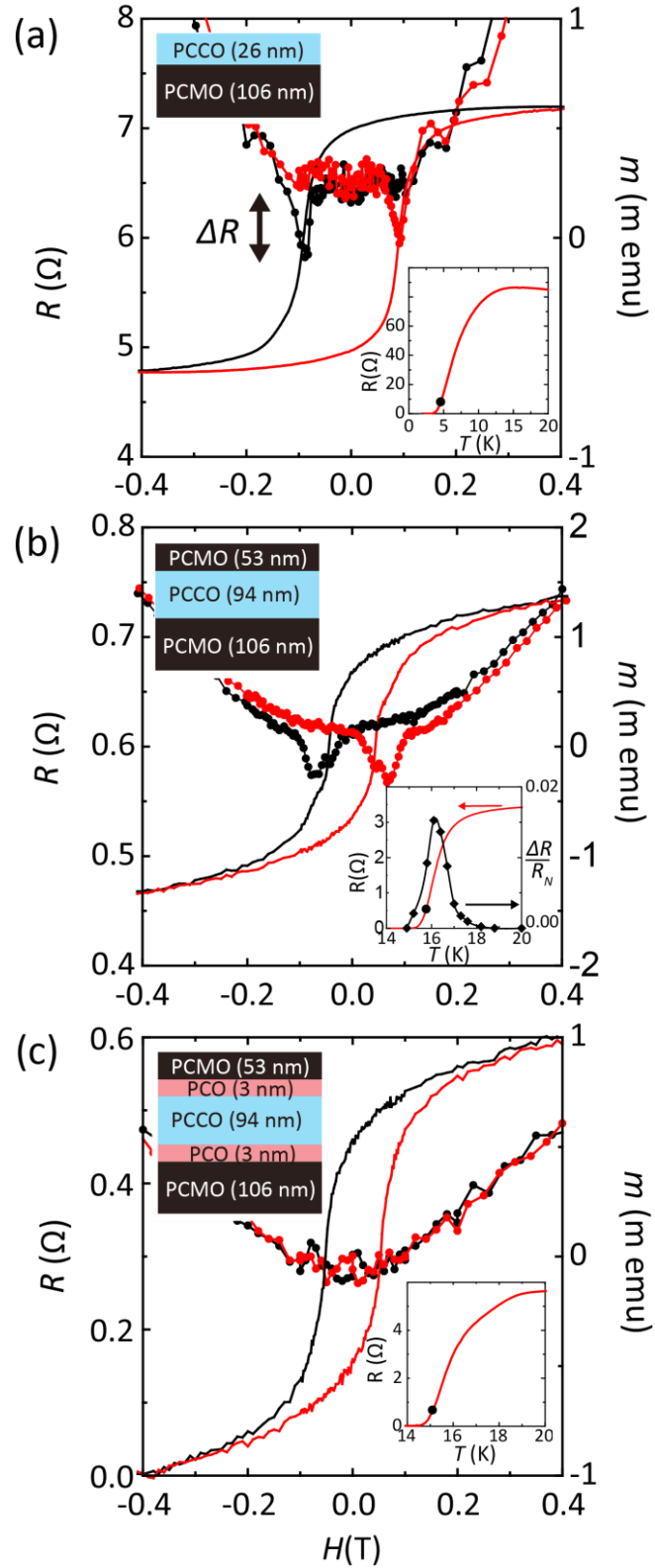

**FIG. 2.** $R(H)$ (left axis) in the superconducting transition and $m(H)$ (right axis) for (a) PCCO(26 nm)/PCMO(106 nm)/STO, (b) PCMO(53 nm)/PCCO(94 nm)/PCMO(106 nm)/STO and (c) PCMO(53 nm)/PCO(3 nm)/PCCO(94 nm)/PCO(3 nm)/PCMO(106 nm)/STO. (a)–(c) (bottom right insets) $R(T)$ with the corresponding temperature for the data in the main panel indicated (round datum point). (b) (bottom right inset) Includes the normalized resistance switching width vs temperature. (Top left inset) Illustrate the corresponding sample structures in (a)–(c).

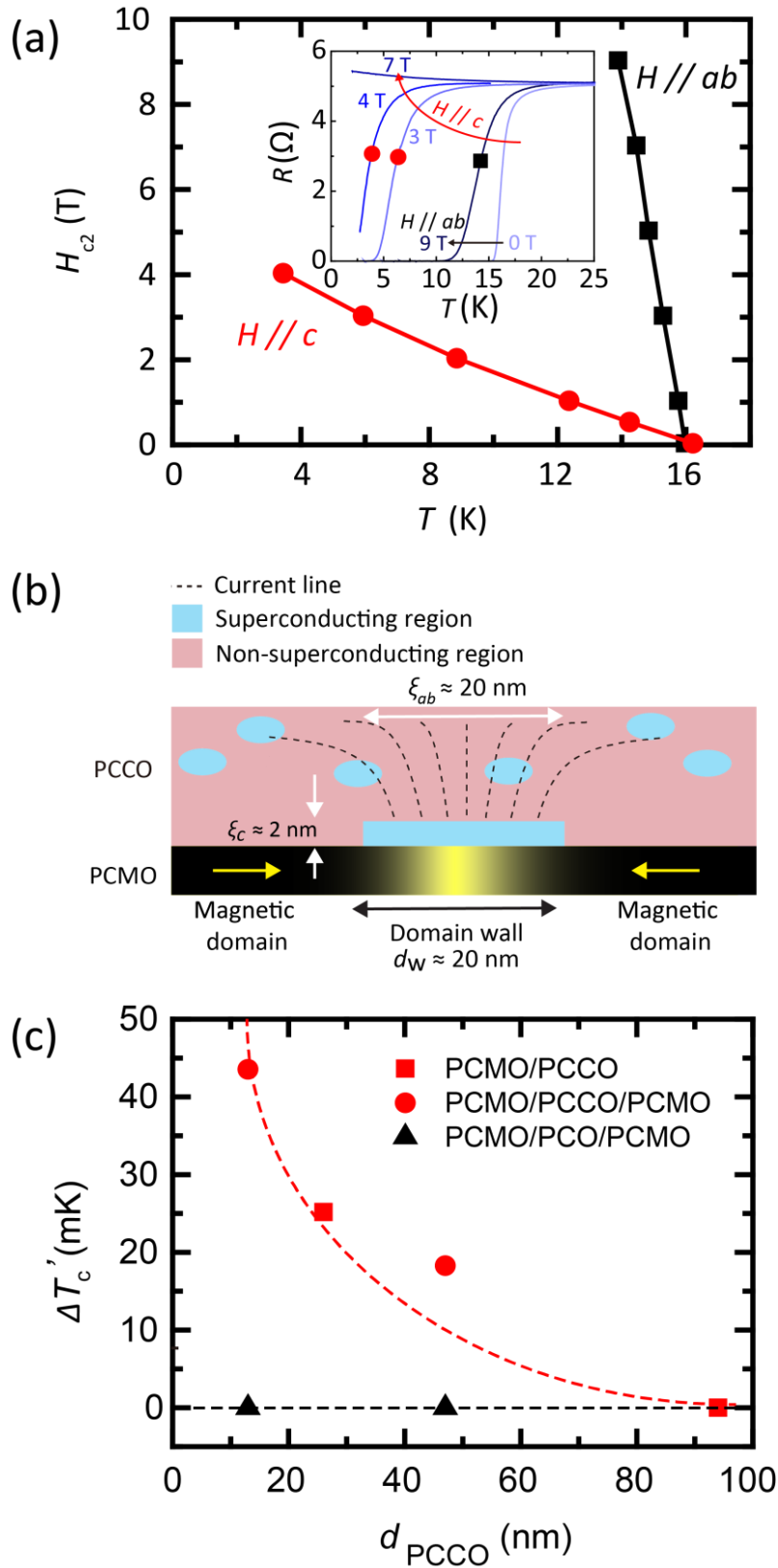

**FIG. 3.** (a) $H_{c2}$ vs $T$ for PCMO(53 nm)/PCCO(94 nm)/PCMO(106 nm). (Inset) $R(T)$ for $H$ approximately parallel to the $ab$ ($H//ab$) and $c$ plane ($H//c$) of PCCO. (b) A schematic illustration of a superconducting film of PCCO on a layer of demagnetized PCMO. (c) The effective $\Delta T_c'$ vs $d_{PCCO}$ for PCMO/PCCO/STO (squares), PCMO/PCCO/PCMO/STO (circles), and PCMO/PCO/PCCO/STO (triangles) samples. Dashed lines are a guide to the eye.

*jjr33@cam.ac.uk

S. Komori,[1] A. Di Bernardo,[1] A.I. Buzdin,[1, 2] M.G. Blamire,[1] and J.W.A. Robinson[1, *]

jjr33@cam.ac.uk

[1] *Department of Materials Science and Metallurgy, University of Cambridge, 27 Charles Babbage Road, Cambridge CB3 0FS, United Kingdom*

[2] *University Bordeaux, LOMA UMR-CNRS 5798, F-33405 Talence Cedex, France*

## 1. Surface morphology of PCMO/PCCO/PCMO/STO

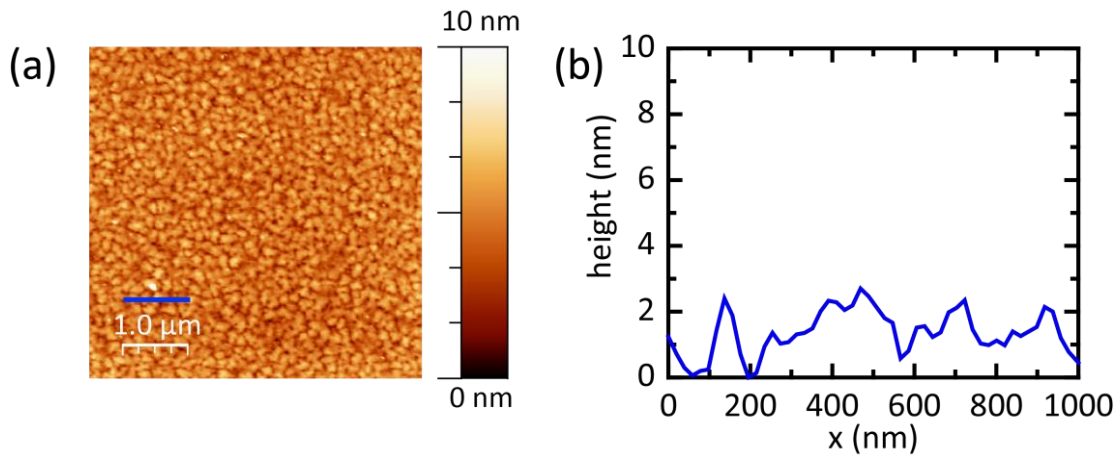

**FIG. S1** (a) Atomic force microscope image on the surface of PCMO(53 nm)/PCCO(26 nm)/PCMO(106 nm)/STO showing a flat surface with a root-mean-square surface roughness of less than 1 nm over an area of 25 $\mu m^2$. (b) Line trace along the blue line in (a).

## 2. Critical temperature and lattice strain in PCCO

We investigated systematically the critical temperature ($T_c$) dependence on PCCO layer thickness ($d_{PCCO}$) for PCCO/STO, PCCO/PCMO/STO, and PCCO/PCO/PCMO/STO. As shown in Fig. S2, PCCO/STO shows a weaker suppression of $T_c$ with decreasing $d_{PCCO}$ compared to PCCO/PCMO/STO and PCCO/PCO/PCMO/STO [Fig. S2(b)]. There are two important factors which influence the $T_c$ of the PCCO thin films in these structures: the concentration of oxygen atoms in the apical positions ($O_a$) of PCCO and interface strain.

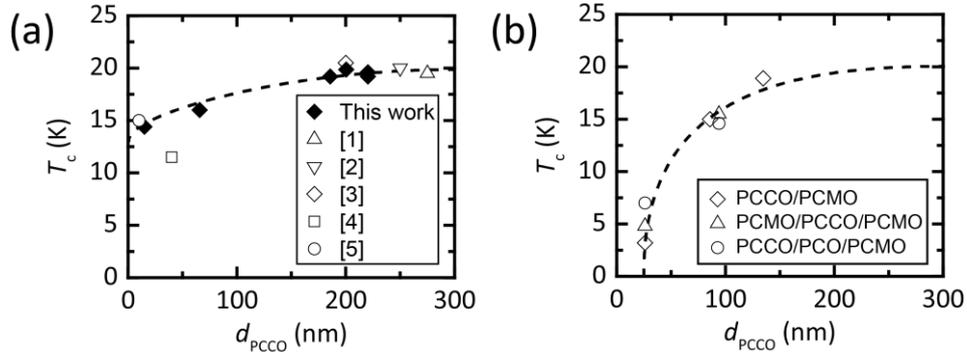

**FIG. S2.** PCCO-thickness-dependence of the critical temperature ($T_c$) for (a) PCCO/STO and (b) PCCO/PCMO/STO, PCMO/PCCO/PCMO/STO and PCCO/PCO/PCMO/STO.

For electron-doped cuprate superconductors a partial removal of $O_a^{2-}$ ions is necessary in order to achieve a (bulk) $T_c$ [6,7]. This is achieved in thin films through the post-annealing treatment of the PCCO following deposition. In the PCCO/PCMO/STO and PCCO/PCO/PCMO/STO structures, the underlying layers of PCMO and PCO will supply additional $O^{2-}$ ions to the PCCO during the annealing process and so reduce the net removal of $O_a^{2-}$ ions from the PCCO near the interface and so suppress the $T_c$ of PCCO on PCMO/STO and PCO/PCMO/STO relative to the $T_c$ of PCCO/STO.

In-plane compressive strain in PCCO will act to enhance the $T_c$ through an extension of Cu-$O_a$ bond length [7-11] and so partially compensate the above effect of annealing. From X-ray reciprocal space maps on the ($\bar{1}09$) diffraction peak of PCCO, we have estimated the PCCO lattice constants in PCCO(23 nm)/STO and PCCO(26 nm)/PCMO(106)/STO (Table S1). From equivalent measurements on a relaxed film of PCCO(200 nm)/STO, we estimate the in-plane strain in PCCO to be larger for PCCO(23 nm)/STO than for PCCO(26 nm)/PCMO(106)/STO as summarized in Table S1. Hence, we expect a lower interface suppression of $T_c$ for the PCCO(23 nm)/STO sample due to greater in-plane strain in combination with enhanced removal of $O_a^{2-}$ ions during the annealing process.

**Table S1:** Lattice parameter data for PCCO on various films determined by reciprocal space maps on ($\bar{1}09$) diffraction peak for PCCO and the ($\bar{1}03$) peak for STO.

| Structure | $d_{PCCO}$ (nm) | $a$ Lattice Constant (nm) | $c$ Lattice Constant (nm) | In-Plane Strain (%) |
|---|---|---|---|---|
| Relaxed PCCO | 200 | 0.3972 | 1.2157 | – |
| PCCO/PCMO/STO | 26 | 0.3969 | 1.2169 | -0.0755 |
| PCCO/STO | 23 | 0.3949 | 1.2173 | -0.5791 |
| STO (substrate) | – | 0.3905 | 0.3905 | – |

### 3. $R(H)$ measurements on YBCO/PCMO/STO and PCCO/LCMO/STO

We investigated the resistance vs in-plane magnetic field properties of YBCO(8 nm)/PCMO(106 nm)/STO and PCCO(28)/LCMO(106)/STO multilayers through the superconducting transition, as shown in Fig. S3.

In contrast to the PCCO/PCMO/STO data in Fig. 2, YBCO(8 nm)/PCMO(106 nm)/STO multilayers do not show switching in $R(H)$ at the coercivity field ($H_c$) of PCMO [Fig. S3(a)] suggesting that the in-plane coherence length [$\xi_{ab}(0)$] of YBCO is too short (~1 nm) to be sensitive to the presence of domain walls in PCMO. In comparison, equivalent $R(H)$ measurements on PCCO(28 nm)/LCMO(106 nm)/STO [Fig. S3(b)] show a negative magnetoresistance in that the resistance through the superconducting transition increases at $H_c$ of LCMO (equivalent to a drop in critical temperature approximately -150 mK). Santamaria et al. reported similar results for LCMO/YBCO/LCMO/STO pseudo spin-valves [12-14] and suggested that the negative switching in $R(H)$ is related to an accumulation of quasiparticle spins in YBCO in the antiparallel state – i.e. when the magnetization directions between the two LCMO layers is antiparallel. For a parallel magnetization alignment between the LCMO layers, spin-polarized quasiparticles from one LCMO layer can easily propagate through YBCO into the second layer of LCMO; however, for an antiparallel alignment, spin-polarized quasiparticles from one layer of LCMO are effectively blocked from entering the second as there are no available states for the quasiparticle spins to enter and so spin accumulation occurs in the central layer of YBCO. This results in a spin-imbalance in YBCO, which acts to suppress the superconducting critical temperature. Although a similar behavior has not been reported for PCCO/LCMO/STO, the same physics may apply: assuming that the spin-diffusion length in PCCO is comparable to the domain wall width in LCMO, at $H_c$ spin-polarized quasiparticles injected into PCCO from one magnetic domain are partially (completely) blocked from entering a neighboring misaligned (oppositely aligned) magnetic domain in LCMO as there are fewer (no) available states for spins to enter. This results in a spin-imbalance in PCCO and therefore to a suppression of superconductivity.

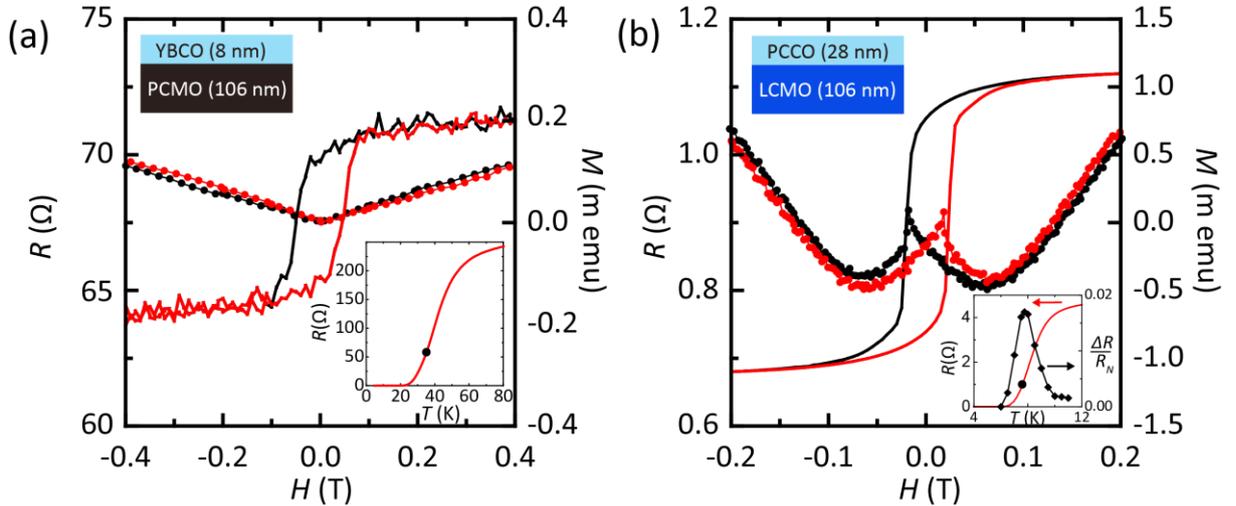

**FIG. S3.** $R(H)$ (left axis) and $M(H)$ (right axis) in the superconducting transition for a (a) YBCO(8 nm)/LCMO(106 nm) bilayer and (b) PCCO(28 nm)/LCMO(106 nm) bilayer. Bottom right insets: $R(T)$ with the measurement temperature in the main panel shown by the black (solid circle) symbols and (b) includes a graph of magnetoresistance through through the superconducting transition vs temperature. Top left inset: schematic illustration of the device structures.